\documentclass[preprint,aps,preprintnumbers,floatfix]{revtex4}

\usepackage{graphicx}
\usepackage{epstopdf}
\usepackage{amsmath}
\usepackage{amssymb}
\usepackage{epsfig}
\usepackage{color}
\usepackage{textcomp}
\usepackage{subfigure}

\begin{document}

\newcommand{\correc}{\bf}
\title{Spin-torque switching and control using chirped AC currents}
\author{Guillaume Klughertz}
\affiliation{Institut de
Physique et Chimie des Mat\'{e}riaux de
Strasbourg, CNRS and Universit\'{e} de
Strasbourg, BP 43, F-67034 Strasbourg, France}
\author{Lazar Friedland}
\affiliation{Racah Institute of Physics, Hebrew University of Jerusalem, Jerusalem 91904, Israel}
\author{Paul-Antoine Hervieux, Giovanni Manfredi}
\email{giovanni.manfredi@ipcms.unistra.fr}
\affiliation{Institut de
Physique et Chimie des Mat\'{e}riaux de
Strasbourg, CNRS and Universit\'{e} de
Strasbourg, BP 43, F-67034 Strasbourg, France.}

\begin{abstract}
We propose to use oscillating spin currents with slowly varying frequency (chirp) to manipulate and control the magnetization dynamics in a nanomagnet. By recasting the Landau-Lifshitz-Slonczewski equation in a quantum-like two-level formalism, we show that a chirped spin current polarized in the direction normal to the anisotropy axis can induce a stable precession of the magnetic moment at any angle (up to $90^\circ$) with respect to the anisotropy axis. The drive current can be modest ($10^6\,\rm A/cm^2$ or lower) provided the chirp rate is sufficiently slow.
The induced precession is stable against thermal noise, even for small nano-objects at room temperature.
Complete reversal of the magnetization can be achieved by adding a small external magnetic field antiparallel to the easy axis. Alternatively,
a combination of chirped ac and dc currents with different polarization directions can also be used to trigger the reversal.
\end{abstract}

\maketitle

\section{Introduction}
Many technological applications of magnetic nano-objects (nanomagnets) require to accurately control their magnetization dynamics \cite{Hillebrands,Back,Gerrits,Schumacher,Seki}.
This can be achieved in several ways, including static or oscillating magnetic fields, thermal effects, and spin-torque transfer (STT). The latter technique consists in injecting a spin-polarized current into a nanomagnet; the electron spins transfer some of their angular momentum to the magnetic material by applying a torque on its magnetic moment and thus inducing the switch.
This technique was first proposed theoretically by Slonczewski \cite{slonczewski} and Berger \cite{berger} and later realized experimentally and further developed by many others \cite{myers,zhang,stiles,katine}.
In the last decade, STT has given rise to new technological
developments such as STT-based random-access memory \cite{wang} and spin-torque nano-oscillators (STNOs) \cite{kiselev}.
Still more recent investigations in this field have been focussing on spin-Hall effects \cite{sinova}.

Achieving optimal switching of the magnetization is a compromise between Joule heating of the sample and reversal time.
Although dc currents are the most widespread method to achieve fast switching \cite{bedau,swiebodzinski}, recent theoretical and experimental work has shown that an ac current tuned at the resonant precession frequency could be even more efficient \cite{cui,chen2011,dunn}.
Various combinations of ac and dc currents and microwave magnetic fields were  implemented to improve the efficiency of the switching \cite{carpentieri,finocchio2006,dunn2014,taniguchi2016}.
A spin current excitation can also be used to induce persistent precession of the magnetic moment, thus enabling magnetic nanostructures to behave as tunable radiofrequency oscillators \cite{houssameddine,bertotti}. Analyzing the tunability and stability of such devices in the presence of intrinsic effects (damping, magnetic anisotropy, thermal fluctuations) is therefore of utmost importance.

In this work, we will demonstrate that an oscillating spin current with slowly variable frequency (chirp) is a very efficient tool for manipulating the magnetization dynamics in a magnetic material. We will focus on two important effects: (i) the fast switching of the magnetic moment and (ii) the precise control of its precession frequency.

A classical nonlinear oscillator can be excited and controlled by a chirped oscillating force using a well-known effect called \emph{autoresonance}, which has been exploited for very diverse applications ranging from plasma \cite{fajans99a}
and atomic \cite{meerson} physics to semiconductor nanostructures \cite%
{manfredi}. Autoresonant excitation occurs when a
nonlinear oscillator starting in equilibrium is driven by a force $F(t)=\epsilon \cos[\int \omega (t)dt]$, with a time-dependent
frequency $\omega (t)$ that slowly passes through the linear frequency $\omega_{0}$ of the oscillator. It can be shown that, for the driving
amplitude $\epsilon $ above a certain threshold $\epsilon _{th}$ (scaling
as $\epsilon_{th}\sim \alpha ^{3/4}$,  where $\alpha =d\omega /dt$ is the chirp rate), the  oscillator frequency \textquotedblleft
locks" to the driving frequency continuously, so that the resonance
condition is preserved for a long time. In that case, the amplitude of the oscillations grows without saturation, until of course some other effects kick in.

In two earlier studies \cite{dyno,2Lautomag}, we made use of the autoresonance mechanism to control the magnetization switching of a magnetic nanoparticle using a chirped microwave field.
This technique was shown to reduce the static switching field and to work well even in the presence of damping, thermal noise, and dipolar interactions.
Here we show that chirped spin currents can be used efficiently to induce the stable precession of the magnetization at a given frequency or to trigger its complete reversal on a nanosecond timescale.

\section{Model and autoresonant excitation}
In the macrospin approximation, the magnetization dynamics is governed by the Landau-Lifshitz-Slonczewski (LLS)
equation \cite{slonczewski,berger}:
\begin{equation}
 \dot{\mathbf{m}} = \boldsymbol\Gamma_{LL} + \boldsymbol\Gamma_{G} + \boldsymbol\Gamma_{ST}, \label{LLG_ST}
\end{equation}
where a dot denotes time differentiation, $\mathbf{m}=\mathbf{M}/\mu_s$ is the normalized magnetic moment of amplitude $\mu_s$, and $\boldsymbol\Gamma_{LL}$, $\boldsymbol\Gamma_{G}$, and $\boldsymbol\Gamma_{ST}$
are the torques induced by the effective magnetic field, the Gilbert damping and the polarized spin current, respectively:
\begin{eqnarray}
\boldsymbol\Gamma_{LL} &=& -\gamma\mu_0 \mathbf{m} \times \mathbf{H}_{\rm eff}, \\
\boldsymbol\Gamma_{G} &=& -\gamma\mu_0 \lambda \mathbf{m} \times (\mathbf{m} \times \mathbf{H}_{\rm eff}),\\
\boldsymbol\Gamma_{ST} &=& -\gamma \mathbf{m} \times (\mathbf{m} \times \mathbf{I}_s),
\end{eqnarray}
where $\gamma=1.76\times 10^{11} \rm rad~ T^{-1}s^{-1}$ is the gyromagnetic ratio, and $\mathbf{I}_s=I_S \mathbf{e}_p$
is the spin current polarized in the direction $\mathbf{e}_p$, expressed in the units of a magnetic field (T).
Here, we neglected the field-like torque term (which is generally small with respect to the spin torque $\Gamma_{ST}$) as well as the angular dependence of the spin torque term, which is also usually small.

The effective field is the sum of an external field and the anisotropy field, $\mathbf{H}_{\rm eff} = \mathbf{H}_0+\mathbf{H}_{an}$. In the present work, we will assume a uniaxial anisotropy along $\mathbf{e}_z$ so that $\mathbf{H}_{an} = \frac{2KV}{\mu_0\mu_s} m_z \mathbf{e}_z$, where $K$ is the anisotropy constant and $V$ the volume. We neglect for the moment the external magnetic field ($\mathbf{H}_0=0$), which will be considered later in Sec. \ref{sec:external}.

Equation (\ref{LLG_ST}) can be rewritten as:
$\dot{\mathbf{m}} = \widetilde{\mathbf{H}} \times \mathbf{m}$, where
\begin{equation}
\widetilde{\mathbf{H}}= \gamma \mu_0\left[\mathbf{H}_{\rm eff} + \lambda(\mathbf{m}\times\mathbf{H}_{\rm eff}) \right]
- \gamma \mathbf{I}_s \times \mathbf{m}. \label{H_tilda}
\end{equation}

We shall adopt an approach due to Feynman \cite{feynman}, which exploits the analogy between the magnetization dynamics
and a two-level quantum-like system and was used earlier to study the autoresonant control of the magnetization dynamics \cite{2Lautomag}. The LLS equation is equivalent to a system of two coupled equations for the complex quantities $A_1$ and $A_2$:
\begin{eqnarray}
i\dot{A_1} &=&\frac{\kappa _{0}}{2}A_{1}+\kappa A_{2}, \label{Feyn1}\\
i\dot{A_2} &=&-\frac{\kappa _{0}}{2}A_{2}+\kappa ^{\ast}A_{1} \label{Feyn2}
\end{eqnarray}
where $\kappa_0 = \widetilde{H}_z$, $\kappa = \frac{1}{2}(\widetilde{H}_x-i\widetilde{H}_y)$, and $\mathbf{m}$ is related to $A_{1,2}$ through the expressions:
\begin{eqnarray}
m_x &=&A_{1}A_{2}^{\ast }+A_{1}^{\ast }A_{2},  \notag \\
m_y &=&i\left( A_{1}A_{2}^{\ast }-A_{1}^{\ast }A_{2}\right) , \\
m_z &=&\left\vert A_{1}\right\vert ^{2}-\left\vert A_{2}\right\vert ^{2}. \notag
\end{eqnarray}
In this formalism, the switching corresponds to a population transfer from, say, level 1 to level 2. Note that $\lvert A_1\rvert^2 + \lvert A_2\rvert^2 = \lvert\mathbf{m}\rvert = 1$, so that the total population (i.e., the total magnetic moment) is conserved.

If we write $A_{1,2}=B_{1,2}e^{i\varphi_{1,2}}$, where $B_{1,2}$ are real functions and $B_1^2 + B_2^2 = 1$, we obtain:
\begin{eqnarray}
m_x &=& 2B_1 B_2 \cos\Delta \varphi,  \notag \\
m_y &=& -2B_1 B_2 \sin\Delta \varphi, \label{mxyz}\\
m_z &=& B_1^2 - B_2^2, \notag
\end{eqnarray}
which shows that, in the Feynman representation, the system is fully described by the real amplitudes $B_{1,2}$ and the phase difference $\Delta \varphi=\varphi_2-\varphi_1$.

In order to illustrate the autoresonant technique, we first consider the simple case where damping is neglected ($\lambda=0$) and the frequency varies linearly with time, $\omega_d(t)=\omega_0 -\alpha t$.
Other effects -- including damping, thermal noise, and an external magnetic field -- will be added in Secs. \ref{sec:precession} and \ref{sec:reversal}.

We focus on the case of a spin current polarized orthogonally to the axis of easy magnetization, i.e., $\gamma \mathbf{I}_s=J_{\perp}(t)\mathbf{e}_x$, with $J_{\perp}(t)=2\varepsilon \cos \varphi_d$ and $\omega_d(t)= \dot{\varphi_d}(t)$ is the chirped driving frequency.
In this case, it follows from Eq. (\ref{H_tilda}) that $\widetilde{\mathbf{H}} = (\omega_r m_z - J_{\perp} m_y) \mathbf{e}_z + J_{\perp} m_z \mathbf{e}_y$, where $\omega_r=2\gamma KV/\mu_s$ is the resonant precession frequency.
The autoresonance mechanism requires that the time-dependent drive frequency crosses the resonant frequency from above, so we set the initial driving frequency $\omega_0 > \omega_r$.

We seek solutions to Eqs. (\ref{Feyn1}) and (\ref{Feyn2}) under the initial conditions $A_1=1$ and $A_2=0$, i.e. $\mathbf{m}=\mathbf{e}_z$.
Using Eqs. (\ref{mxyz}), we obtain:
\begin{eqnarray}
\dot{B_1} &=& -(J_{\perp}/2) (B_1^2-B_2^2)B_2\cos\Delta \varphi \label{1}\\
\dot{B_2} &=& (J_{\perp}/2) (B_1^2-B_2^2)B_1\cos\Delta \varphi \label{2}\\
\Delta\dot{\varphi} &=& -\omega_r (1-2B_2^2) + J_{\perp} (B_2/B_1) \sin\Delta \varphi\label{3}.
\end{eqnarray}
We then define $\phi=\Delta\varphi-\varphi_d-\pi/2$, and use the rotating wave approximation (neglecting the high frequencies)
to derive the equations for the coupled variables $B_2$ and $\phi$:
\begin{eqnarray}
\dot{B_2}  &=& -(\varepsilon/2) (B_1^2-B_2^2)B_1\sin\phi \label{B2eq}\\
\dot{\phi}  &=& \omega_r - \omega_d-2\omega_r B_2^2  -\varepsilon/(2B_1B_2)\cos\phi, \label{phieq}
\end{eqnarray}
where we recall that $B_1=\sqrt{1-B_2^2}$.
Focussing on the weakly nonlinear regime ($B_1\approx1$ and $B_2 \ll 1$), we obtain:
\begin{eqnarray}
\dot{B_2}  &=& -(\varepsilon/2) \sin\phi\\
\dot{\phi} &=&  \omega_r - \omega_d -2\omega_r B_2^2 -(\varepsilon/2B_2)\cos\phi.
\end{eqnarray}
The above equations are typical of systems that can be driven into autoresonance \cite{2Lautomag}. Previous work \cite{Fajans} showed that the system is captured into autoresonance when the excitation amplitude $\varepsilon$ exceeds a certain threshold
\begin{equation}
\varepsilon>\varepsilon_{th}=0.82 (2\omega_r)^{-1/2} \alpha^{3/4}. \label{threshold}
\end{equation}

\begin{figure}[h!]
\centering
\includegraphics[width=0.8\linewidth]{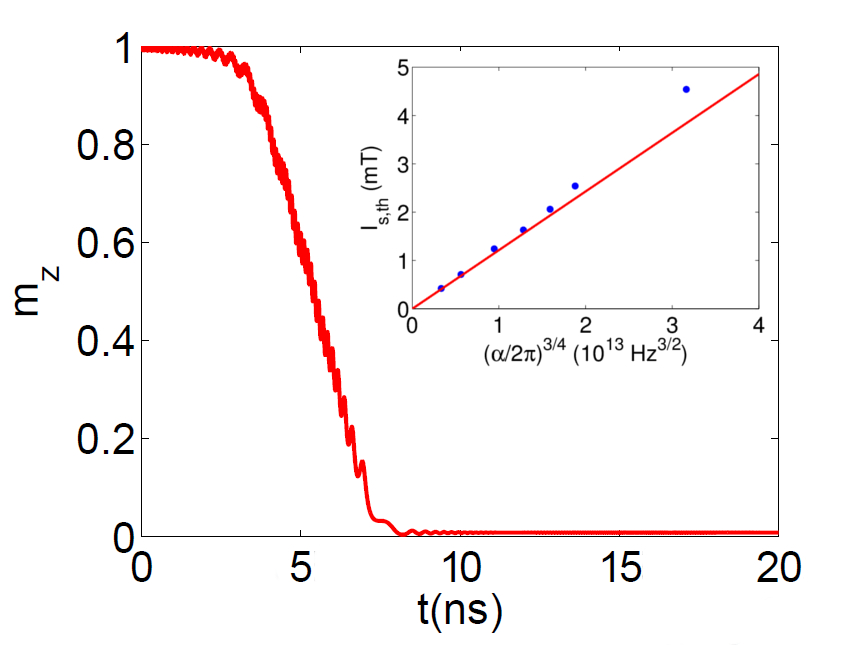}
\caption{Dynamics of $m_z$ for a polarized spin current with amplitude $\varepsilon=3 \varepsilon_{th}$, initial frequency $\omega_0/2\pi=20$~GHz, and linear chirp rate $\alpha=2$~GHz/ns. The inset shows the threshold amplitude $I_{s,th}$ against $\alpha^{3/4}$. The blue dots are numerical results obtained by solving the full LLS equation, while the red line represents the theoretical formula, Eq.(\ref{threshold}).}
\label{fig:Jperp_auto}
\end{figure}

When the above condition is satisfied, the chirped spin current stays locked with the precession oscillations, and drives the magnetic moment away from the anisotropy axis even in the nonlinear regime.
These theoretical results are in agreement with numerical simulations of the full Landau-Lifshitz-Slonczewski equation, carried out for a nanomagnet with volume $V=2\times 10^{-24}\rm m^3$ ($20\rm nm\times 20 \rm nm\times 5 \rm nm$), anisotropy constant $K=2.2\times10^5$ J/m$^3$, and magnetic moment $\mu_s=3.35\times10^{-18}$ J/T (see Fig. \ref{fig:Jperp_auto}).
For these parameters, the resonant frequency is $\omega_r /2\pi =7.36$~GHz.

Note that, according to Eq. (\ref{B2eq}), the time derivative of $B_2$ vanishes when $B_1=B_2$. Thus, when $m_z=0$, it is impossible to further populate the level $B_2$. This implies that one cannot fully reverse the magnetization (i.e., reach $m_z=-1$) using such spin current. The largest precession angle attainable with this technique is $\theta =90^\circ$ (where $\theta$ is the angle between the magnetic moment and the $z$ axis) as can be seen from Fig. \ref{fig:Jperp_auto}.
In the absence of damping and thermal noise, the magnetic moment will precess indefinitely perpendicular to the anisotropy axis $\mathbf{e}_z$.

However, we will show in Sec. \ref{sec:external} that, by adding a small ($\approx 10\,\rm mT$) external magnetic field antiparallel to the anisotropy axis, it is possible to fully reverse the magnetization using the autoresonant technique described above.
A second reversal technique, based on the combination of {\em two} spin currents, parallel and perpendicular to $\mathbf{e}_z$, will be illustrated in Sec. \ref{sec:parallel}.

\section{Autoresonant control of the precession}\label{sec:precession}

We now show that the autoresonant technique can be used to bring the magnetic moment to rotate around the anisotropy axis at a certain target angle and precession  frequency. This is an important feature that allows to convert an electric current into high-frequency magnetic rotation, with potential applications to nanoscale devices such as STNOs. In particular, we want to study the stability of the forced precession regime using a spin current, including the effect of the Gilbert damping term $\boldsymbol\Gamma_{G}$ and thermal fluctuations.

To this end, we enforce a fixed precession angle by chirping the excitation frequency exponentially, from the initial value $\omega_0$ towards the asymptotic value $\omega_f$:
\[
\omega_d = \dot{\varphi_d}= \omega_f + (\omega_0-\omega_f) \mathrm{e}^{-t/\tau}.
\]
However, we emphasize that the particular form of the function  $\omega_d(t)$ is not important -- the autoresonant mechanism works  in any case as long as the frequency variation is sufficiently slow.

\subsection{Gilbert damping and stability properties}
We proceed from Eqs. (\ref{B2eq})-(\ref{phieq}), where we add a small dissipative term ($\lambda \omega_r/\varepsilon \ll 1$). Assuming that, for $\omega_d < \omega_r$ (i.e., after crossing the linear resonance), the system is sufficiently excited so that $B_2$ is finite and $\epsilon/(B_1 B_2)\ll 1$, we can neglect the $\cos\phi$ term in Eq. (\ref{phieq}). Then we get:
\begin{eqnarray}
\dot{x} &=& -\varepsilon F(x)\sin\phi -2\lambda\omega_r G(x) \label{xeq_ST}\\
\dot{\phi} &=&\Delta- 2\omega_r x, ~~~~~\Delta>0\label{phieq_ST}
\end{eqnarray}
where $x=B_2^2$, $\Delta=\omega_r-\omega_d$, $F(x)=(1-2x)\sqrt{x(1-x)}$ and $G(x)=(1-2x)x(1-x)$.
The steady state of this system is $x_0=\Delta/(2\omega_r)$, $\phi_0=\pi+\frac{2\omega_r\lambda}{\varepsilon} \frac{G_0}{F_0}$, where
$F_0=F(x_0)$ and $G_0=G(x_0)$.
We now discuss the stability of this steady state with respect to small perturbations, by writing $x=x_0+\delta x \mathrm{e}^{i\nu t}$ and $\phi=\phi_0+\delta \phi\mathrm{e}^{i\nu t}$. Equations (\ref{xeq_ST}) and (\ref{phieq_ST})
lead to the characteristic equation $\nu^2-2i\lambda \omega_r f_0\nu-2\varepsilon \omega_r F_0=0$,
where $f_0 \equiv G_0' -F_0'(G_0/F_0)=(1-2x_0)^2/2$ (the apex denotes differentiation with respect to $x$), yielding two characteristic frequencies
\begin{equation}
\nu_{\pm}=i\lambda\omega_r f_0 \pm \sqrt{2\varepsilon\omega_r F_0 - (\lambda\omega_r f_0)^2}. \label{eq:charact}
\end{equation}
As the last term in the square root is small and $f_0$ is positive, both roots $\nu_{\pm}$ have a positive imaginary part, which guarantees stability. Thus, the autoresonant regime is always stable, despite the fact that the damping tends to bring the magnetic moment back to the anisotropy axis.

\begin{figure}
\centering
{\includegraphics[width=0.85\linewidth]{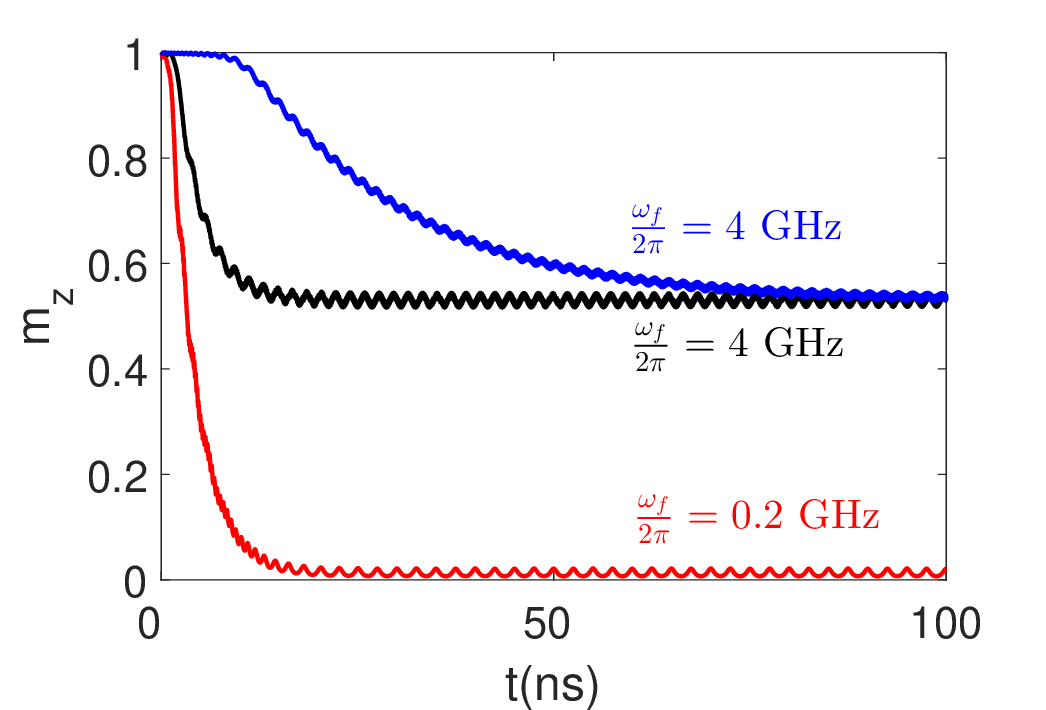}}
\caption{Time evolution of $m_z$ for a polarized chirped spin current with initial frequency $\omega_0/2\pi = 10$GHz. The final frequencies and currents are: $\omega_f/2\pi = 4$~GHz, $I_S=6.3$ mT (black and blue curves)
and $\omega_f/2\pi = 0.2$~GHz, $I_S=11.3$ mT (red curve). The chirp time is $\tau=2.5~\rm ns$ for the 0.2~GHz case. For the 4~GHz cases, we used $\tau=2.5~\rm ns$ (black curve) and $\tau=7.5~\rm ns$ (blue curve).}
\label{fig:mz_stable_ST}
\end{figure}

We have checked numerically, by solving the full LLS equation, that stable precession of the magnetic moment can indeed be forced for any angle in the range $[0,\pi/2]$ using the autoresonance technique. Some examples are shown in Fig. \ref{fig:mz_stable_ST}, for final frequencies $\omega_f/2\pi  = 4$~GHz and 0.2~GHz, which correspond respectively to angles $\theta=57^\circ$ and $88^\circ$ between the magnetic moment $\mathbf{m}$ and the anisotropy axis $\mathbf{e}_z$.
In the same figure, we also show the effect of the chirp time $\tau$. The latter can be used to control precisely the magnetization dynamics, so that the magnetic moment reaches its final precession orbit with the desired speed. For instance, in Fig. \ref{fig:mz_stable_ST}, two cases are shown for $\omega_f/2\pi = 4$~GHz with the asymptotic precession being achieved in either $\sim 20~\rm ns$ or 80~ns.

\begin{figure}
\centering
\subfigure{\includegraphics[width=0.65\linewidth]{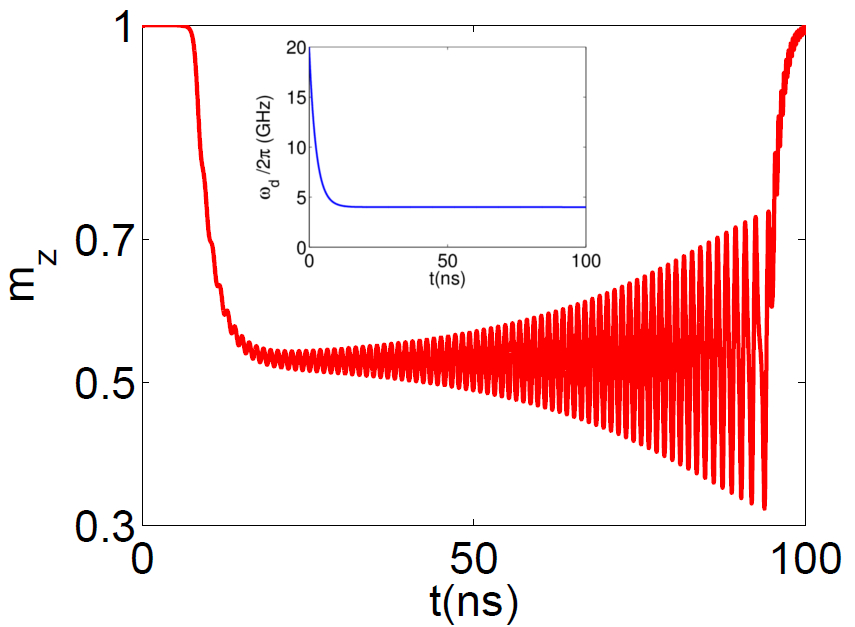}}\\
\subfigure{\includegraphics[width=0.65\linewidth]{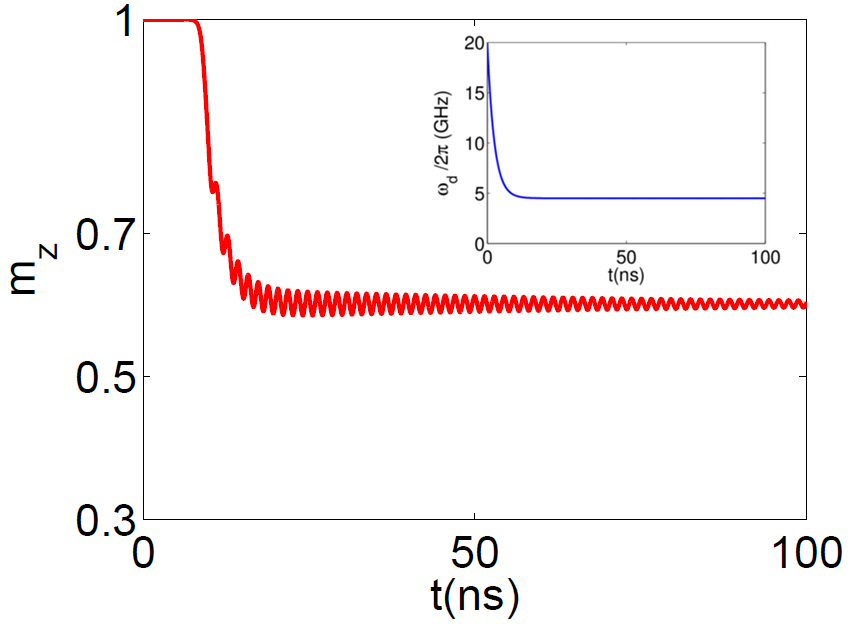}}
\caption{Evolution of $m_z$ for a chirped magnetic field excitation, for an unstable case with $\omega_f^\infty/2\pi=4$~GHz (top frame) and a stable case with $\omega_f^\infty/2\pi=4.5$~GHz (bottom frame). The insets show the temporal profile of the drive frequency $\omega_d(t)$.}
\label{fig:mz_stable_MW}
\end{figure}

In contrast, when the magnetization dynamics is excited with a chirped oscillating magnetic field (usually in the microwave range\cite{dyno}), a similar analysis yields instability for $\theta>45^\circ$. Numerical simulations of the full Landau-Lifshitz-Gilbert equation, similar to those we performed in an earlier work \cite{dyno}, confirm this result, as can be seen from Fig. \ref{fig:mz_stable_MW}. It is observed that the stability threshold is around $\theta^\infty \approx 50^\circ$, slightly larger than the theoretical value.

An important advantage of the autoresonant drive is that the ac current can be arbitrarily small provided the chirp rate is slow enough, as is apparent from the threshold condition, Eq. (\ref{threshold}). For instance, for the nano-magnets considered in the preceding section, a current density of 3~mT in magnetic field units corresponds
\footnote{The conversion is: $j[{\rm A/m^2}] = j[{\rm T}] \times (2e\mu_s d)/(V\hbar)$, where $d$ is the thickness of the magnetic layer.}
to roughly $7\times 10^6\,\rm A/cm^{2}$, which is a standard value for STNOs \cite{zeng}.
For this current, the threshold chirp time $\alpha^{-1/2}$ is of the order of 0.5~ns (the actual time to reach the asymptotic precession angle will be a multiple of this time), as can be deduced from the inset of Fig. \ref{fig:Jperp_auto}.
But since the threshold current decreases almost linearly with decreasing $\alpha$, using a slower chirp can reduce the required current by a significant factor. For instance, decreasing $\alpha$ by a factor of 10, cuts the threshold current by a factor $10^{3/4} \approx 5.6$, while it increases the time to induce the precession by a factor $10^{1/2} \approx 3.2$.
Of course, there is a trade-off to be made between the rapidity of the overall process and the intensity of the required current, but it is clear that competitively low currents can be achieved if one accepts to lengthen the time to induce the precession.

\subsection{Thermal effects}
In the results reported above, temperature effects were neglected. However, previous theoretical \cite{dyno,barth} and experimental \cite{shalibo} studies showed that the autoresonant mechanism is rather robust against thermal noise. In order to check that the same conclusion holds in the present case, we introduced thermal fluctuations in our model. As is usually done \cite{dyno}, thermal fluctuations at temperature $T$ are represented as a random magnetic field
$\tilde{\mathbf b}(t)$ with zero mean and autocorrelation function given by:
\begin{equation}
\langle \tilde{b}_i (t) \tilde{b}_j (t') \rangle = \frac {2 \lambda k_B T} {(1+ \lambda^2) \gamma \mu_s} \delta_{ij} \delta (t-t'),
\end{equation}
where $i,j$ denote the cartesian components $(x,y,z)$, $\delta_{ij}$ is the Kronecker symbol (meaning that the spatial components of the random field are uncorrelated),
and $\delta (t-t')$ is the Dirac delta function, implying that the autocorrelation time of $\tilde{\mathbf b}$ is much shorter than the response time of the system. The temperature is thus proportional to the autocorrelation function of the fluctuating field.

\begin{figure}
\centering
{\includegraphics[width=0.8\linewidth]{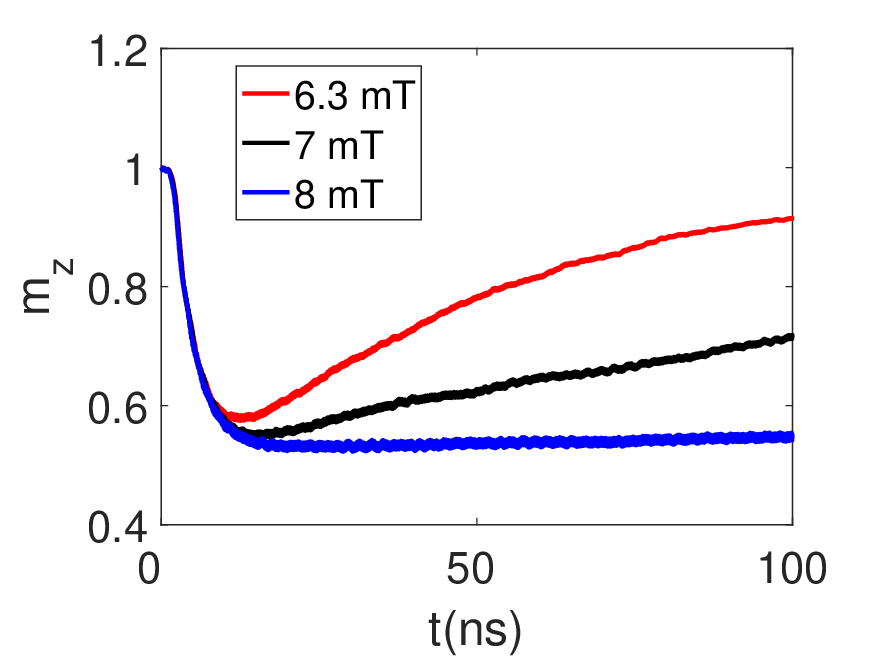}}
\caption{Finite temperature effects ($T=300\,\rm K$): Time evolution of $m_z$ for a polarized chirped spin current with initial frequency $\omega_0/2\pi = 10$GHz and final frequency $\omega_f=4\rm ~GHz$, for three values of the spin current: $I_S=6.3$ mT (red curve, theoretical threshold at $T=0$), $I_S=7$ mT (black), and $I_S=8$ mT (blue).}
\label{fig:temperature}
\end{figure}

In Fig. \ref{fig:temperature}, we plot results at room temperature ($T=300 \rm K$) for a 25nm-diameter nanoparticle (blocking temperature $\sim 5000 ~\rm K$) with damping $\lambda=0.01$ and $\omega_f/2\pi=4\rm ~GHz$. There is no external magnetic field. The three curves correspond to different values of the oscillating spin current amplitude. The amplitude $I_S = 6.3 \rm mT$ is just above the autoresonant threshold in the absence of thermal fluctuations and can thus control the precession in a stable way, as was done in Fig. \ref{fig:mz_stable_ST} (black curve). However, this is no longer true at finite temperature (Fig. \ref{fig:temperature}), where thermal noise drives the magnetic moment back to the $z$ axis. In order to induce a stable precession, the current needs to be increased slightly, up to 8~mT or higher.

The above phenomenon is consistent with what was observed in the past for finite-temperature systems that are excited autoresonantly \cite{dyno,barth,shalibo}.
In particular, the ability to hold the precession for increasing driving amplitude $I_S$ (Fig. \ref{fig:temperature}) can be explained as follows. The autoresonant system is formally equivalent to a quasiparticle trapped in an effective potential well of height $V_0$ proportional to $I_S$ \cite{barth}. The noise drives the quasiparticle out of the well, on a time scale proportional to $\exp(V_0/k_B T)$ if the quasiparticle is initially deeply trapped in the well \cite{dykman}. Therefore, increasing $I_S$ (and thus $V_0$) amounts to reducing the effect of the thermal noise, in accordance with Fig. \ref{fig:temperature}.
In addition, thermal fluctuations also modify the threshold phenomenon. At zero temperature, there exists a sharp threshold for the excitation amplitude $I_S$ above which the system is always captured into the autoresonant regime. In the present work the existence of such a threshold, which depends on the chirp rate $\alpha$, was confirmed in Fig. \ref{fig:Jperp_auto} (see inset). At finite temperature, the threshold is no longer sharp, but instead displays a certain width that is proportional to the square root of the temperature \cite{dyno}.
All these effects were observed in our numerical simulations in full agreement with the general autoresonance theory.

The above results show that the autoresonant technique is very stable against thermal fluctuations. Such stability properties are of great importance in real STT devices \cite{kim}, where phase fluctuations due to the presence of thermal noise can have a disruptive effect. Here, we showed thermal fluctuations do not disrupt the autoresonant drive of the precession, provided the spin current is increased slightly above the nominal (zero-temperature) threshold. In addition, the autoresonant excitation is not sensitive to the precise temporal profile of the chirped current frequency, the only requirement being that the frequency varies slowly in time.

We also note that many simulations of STNOs were performed at zero \cite{Taniguchi2014} or very small \cite{Rippard2010} temperature, or they involved large nano-objects \cite{Kubota2013} (diameter $> 100$nm) for which the blocking temperature is very high and therefore the effect of thermal noise is minor even at $T=300$K.
The present autoresonant technique has proven to preserve the stability of the oscillations even for much smaller nano-objects (25nm) at room temperature. It may therefore be more advantageous for such ultrasmall nano-oscillators.

\subsection{Phase locking}
The standard way to induce a precession at a given frequency is to use a dc spin current, which counteracts the Gilbert damping term, thus preventing the magnetic moment to relax back to easy axis \cite{slavin,Rippard2010,Zeng2012,Taniguchi2014}.
Although a dc current may be easier to implement, our approach has some specific advantages. First, it is possible (by modulating the frequency variation) to control precisely the trajectory of the magnetic moment towards the desired precession angle. Second, the method is rather stable against damping and thermal fluctuations, as was shown in the preceding paragraphs.

Now, we show that the autoresonant technique is also useful to induce {\em phase locking} between the external signal and the response of the STNO. Usually, phase locking (or injection locking) is achieved by combining an external dc current with an ac drive signal \cite{Rippard}. When the ac drive is close enough to the natural frequency of the STNO, then the latter starts oscillating in phase at the same frequency of the drive. For a given dc current, phase locking is achieved only for a narrow range of drive frequencies.

\begin{figure}
\centering
\subfigure{\includegraphics[width=0.7\linewidth]{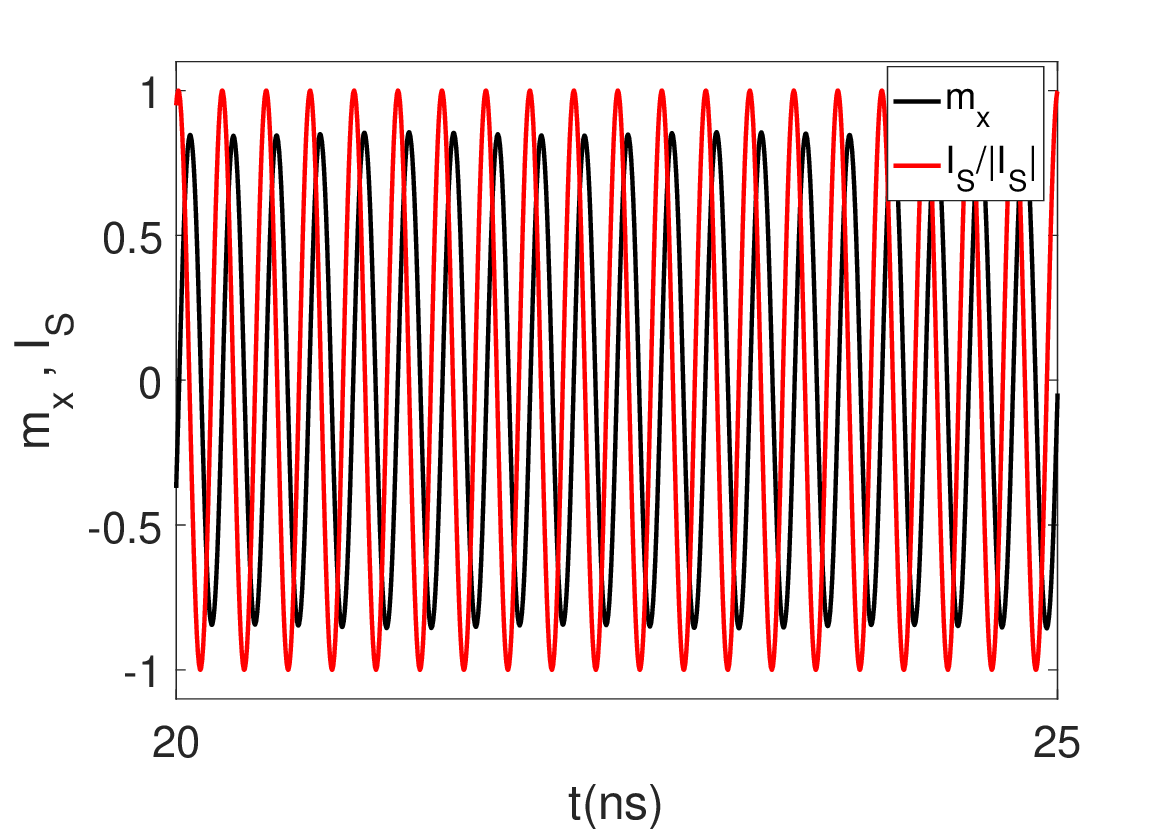}}\\
\subfigure{\includegraphics[width=0.7\linewidth]{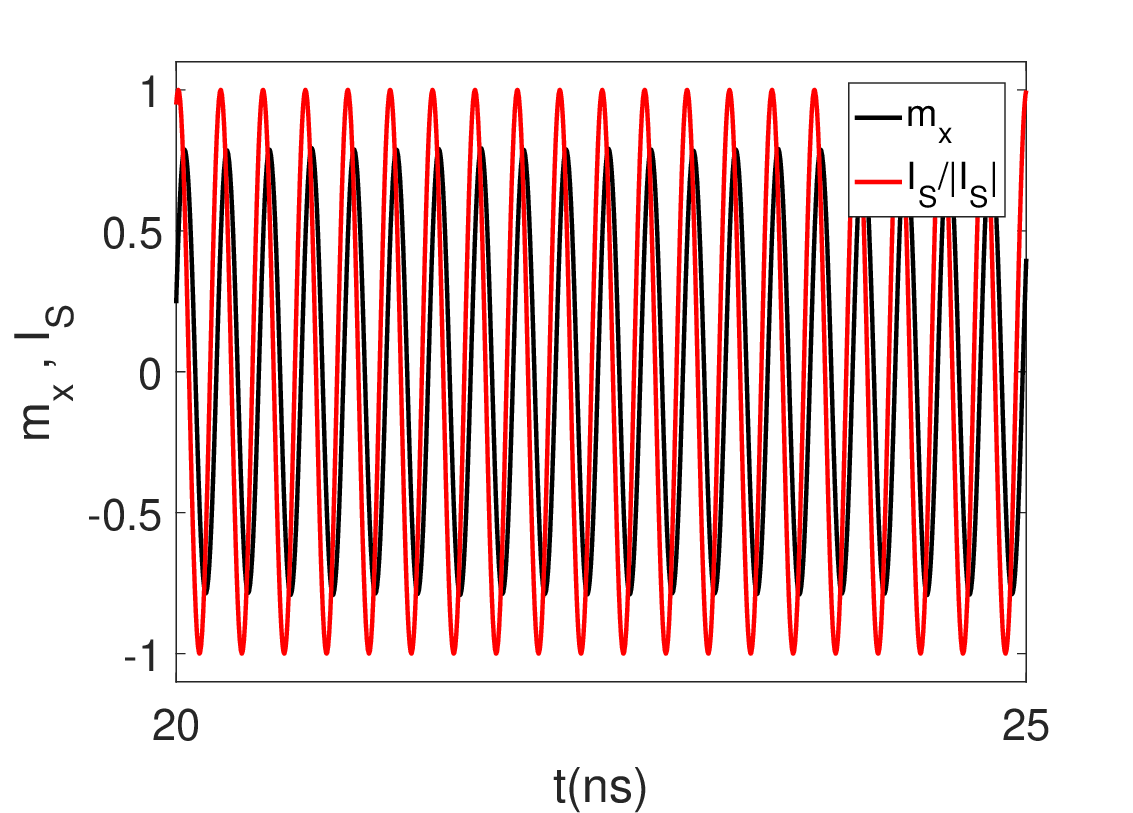}}
\caption{Time-domain evolution of the driving spin current normalized to its maximum value (red curve) and the $m_x$ component of the magnetic moment (black curve). The top panel corresponds to the same parameters as the black curve in Fig. \ref{fig:mz_stable_ST}, at zero temperature. The bottom panel includes thermal fluctuations at $T=300\,\rm K$ and the drive current is slightly larger, $I_S=8\,\rm mT$.}
\label{fig:phaselock}
\end{figure}

Using our approach, it was possible to phase-lock the drive (chirped ac current) to the STNO precession response, without any external dc currents and for a wide range of precession frequencies. Indeed, the autoresonant technique was originally deviced exactly for such a purpose: to bring a system to oscillate at a specified nonlinear frequency by slowly sweeping the frequency of the drive. This should work for any target frequency, provided the threshold condition, Eq. (\ref{threshold}), is satisfied. Importantly, the threshold condition also tells us that the driving ac current can have a very small amplitude, provided the frequency variation rate is slow enough.

To demonstrate phase locking between the drive and the STNO precession, we plot in Fig. \ref{fig:phaselock} (top) the driving spin current and the $x$ component of the magnetic moment for the same case as the black curve ($\omega_f = 2\pi \times 4\,\rm GHz$) in Fig. \ref{fig:mz_stable_ST}. It is clear that the two quantities evolve in phase, and they stay so for very long times (we only show a limited time window for clarity). Importantly, the phase locking appears to be robust against thermal fluctuations, as is shown in Fig. \ref{fig:phaselock} (bottom). Such robustness and flexibility should make the proposed technique competitive with respect to other approaches.

\section{Magnetization reversal}\label{sec:reversal}
As a further application, we propose two procedures to completely switch the magnetic moment from parallel to antiparallel to the anisotropy axis $\mathbf{e}_z$.
The first procedure is based on an external static magnetic field antiparallel to the anisotropy axis, combined with the autoresonant spin current described in the preceding sections.
The second method relies on the combination of two types of spin currents (ac and dc) polarized in different directions.

\subsection{External magnetic field}\label{sec:external}
\begin{figure}
\centering
{\includegraphics[width=0.8\linewidth]{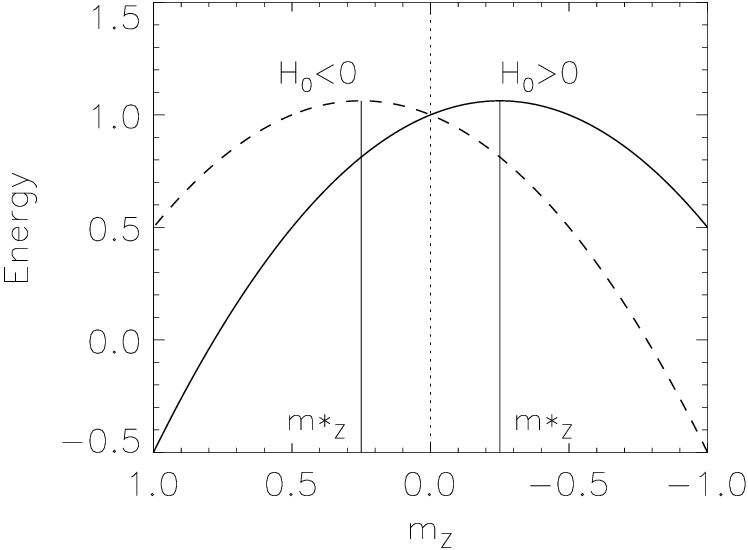}}
\caption{Schematic view of the energy barrier as a function of $m_z$ for two cases with external field parallel ($H_0>0$) or antiparallel ($H_0<0$) to the $z$ axis. $m_z^\star$ denotes the peak of the barrier in either case. The point $m_z=0$ cannot be crossed through autoresonant excitation.}
\label{fig:barrier}
\end{figure}

The presence of an external magnetic field $\mathbf{H}_0=H_0 {\mathbf{e}}_z$ affects the magnetization dynamics in two ways, through the torques $\Gamma_{LL}$ and $\Gamma_{G}$. As to $\Gamma_{LL}$, its  primary effect is to move the peak of the energy barrier (the point where the instantaneous precession frequency vanishes) away from $\theta = 90^\circ$ (i.e., $m_z=0$), towards values $\theta < 90^\circ$ ($m_z>0$) for an external field antiparallel to ${\mathbf{e}}_z$, and $\theta > 90^\circ$ ($m_z<0$) for a field parallel to ${\mathbf{e}}_z$ (Fig. \ref{fig:barrier}).
As to $\boldsymbol\Gamma_{G}$, the part of the Gilbert torque that is due to the external field can be written:
\[
\boldsymbol\Gamma_{G}^{ext} = -\gamma\mu_0 \lambda \mathbf{m} \times (\mathbf{m} \times \mathbf{H}_0) = -\gamma \mu_0\lambda H_0 (m_z \mathbf{m}- {\mathbf{e}}_z).
\]
Therefore, the $z$ component of the magnetic moment evolves under the action of  $\boldsymbol\Gamma_{G}^{ext}$ as follows:
\begin{equation}
\dot{m}_z = \gamma\mu_0 \lambda H_0 (1-m_z^2) + \dots
\label{gtorqueext}
\end{equation}
Of course, many other terms (notably the spin current) also affect the evolution of $m_z$.
From the above expression, we see that the effect of $\boldsymbol\Gamma_{G}^{ext}$ is to drive the magnetic moment towards $m_z=-1$ when $H_0<0$ and towards $m_z=1$ when $H_0>0$.

However, according to Eq. (\ref{B2eq}), the autoresonant condition is still lost at $\theta = 90^\circ$ (when $B_1 = B_2$, or $m_z=0$), irrespective of the external field.
Thus, we have two possible scenarios, depending on the orientation of the external field (see Fig. \ref{fig:barrier}):
\begin{enumerate}
\item
If $H_0<0$ (antiparallel) the peak of the energy barrier is situated at a position $1>m_z^\star>0$. Starting from $m_z=1$, the autoresonant excitation induces precession with decreasing $m_z$ and can bring the magnetic moment to overcome the energy barrier. Subsequently, the autoresonant phase locking is lost and the external-field Gilbert torque $\boldsymbol\Gamma_{G}^{ext}$ drives the magnetic moment towards $m_z=-1$.
\item
If $H_0>0$ (parallel) the peak of the energy barrier is situated at a position $m_z^\star<0$. The autoresonant excitation can never bring the magnetic moment to cross the $m_z=0$ plane and thus it can never overcome the barrier. In this case, $\boldsymbol\Gamma_{G}^{ext}$ brings the magnetic moment back to to its initial value $m_z=1$ [see Eq. (\ref{gtorqueext})].
\end{enumerate}

\begin{figure}
\centering
{\includegraphics[width=0.8\linewidth]{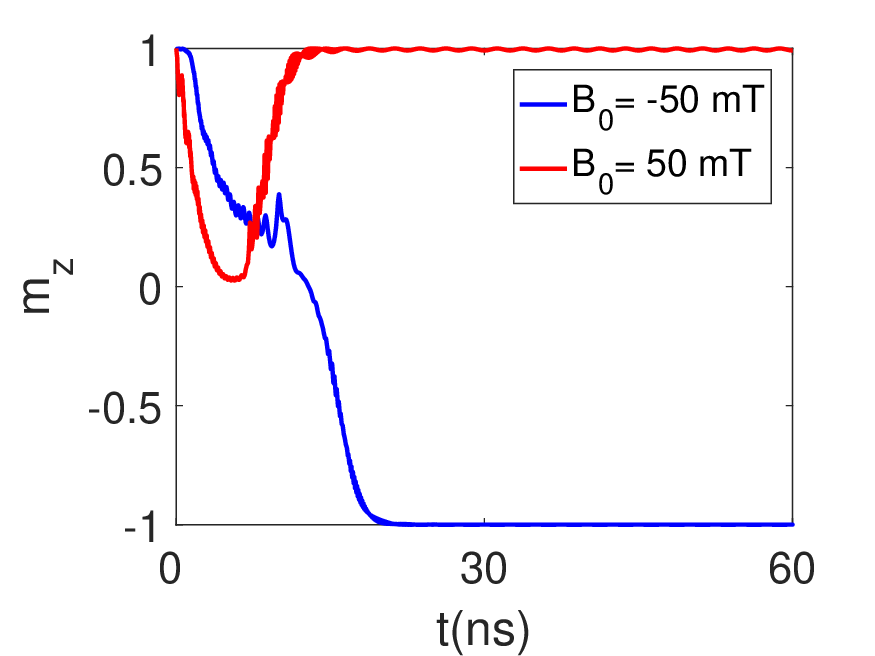}}
\caption{Evolution of the $m_z$ component of the magnetic moment for a case with external magnetic field parallel ($\mu_0 H_0=50\,\rm mT$, red curve) or antiparallel ($\mu_0 H_0=-50\,\rm mT$, blue curve) to the $z$ axis. The driving spin current is $I_S=11.3\,\rm mT$ for the parallel case and $I_S=40\,\rm mT$ for the antiparallel case.}
\label{fig:external}
\end{figure}

In Fig. \ref{fig:external}, we present some numerical results that confirm the above scenarios. We consider an external field of intensity $H_0 = \pm 50 \,\rm mT$, oriented either parallel or antiparallel to the anisotropy axis $z$. Other parameters are identical to those corresponding to the red curve on Fig. \ref{fig:mz_stable_ST}.
When the magnetic field is antiparallel to $\mathbf{e}_z$, the magnetic moment first starts precessing at increasing azimuthal angle until it crosses the barrier, which is located around $\theta=79^\circ$ ($m_z^\star=0.19$, visible on Fig. \ref{fig:external} as the point where the autoresonant phase locking is lost). Subsequently, the magnetic moment relaxes towards $m_z = -1$ under the action of the external-field torque.
In contrast, when $\mathbf{H}_0$ is parallel to $\mathbf{e}_z$, the  magnetic moment goes back to its original position $m_z = +1$, in agreement with the second scenario of our analysis.

We note that for $H_0<0$ we were able to reverse the magnetic moment, in contrast to the case with no external field, for which the plane $m_z=0$ could not be crossed. Thus, adding a small antiparallel magnetic field seems to be a good strategy to obtain complete reversal of the magnetization on a nanosecond timescale using the proposed autoresonant technique. The initial stage of the reversal (up to the top of the energy barrier) is driven by the autoresonant excitation; then, once the barrier has been passed, the Gilbert damping term brings the magnetization to the opposite direction. In this case, the plane $m_z=0$ is not crossed during the autoresonant stage of the evolution, so that the restriction of Eq. (\ref{gtorqueext}) does not apply.
Note that a very small antiparallel magnetic field is sufficient to trigger the complete reversal: for instance, it works fine for $H_0=-10\,\rm mT$, for which the energy barrier is situated at $\theta=88^\circ$ (not shown here).

\subsection{Parallel spin current}\label{sec:parallel}

The procedure is again based on the autoresonance technique and requires {\em two} spin currents polarized in the parallel and perpendicular directions with respect to $\mathbf{e}_z$.
Let us first consider a purely parallel spin current: $\gamma \mathbf{I}_s=-J_{\parallel}(t)\mathbf{e}_z$.
The effective field is then given by (we neglect damping for simplicity):
\[
\widetilde{\mathbf{H}}= \omega_r m_z \mathbf{e}_z - J_{\parallel}( m_y \mathbf{e}_x- m_x \mathbf{e}_y).
\]
Using the two-level formalism described above, one can derive a closed-form solution for the real amplitude $B_2$:
\begin{equation}
B_2^2(t) = \frac{B_2^2(0)\mathrm{e}^{2\Gamma}}{B_1^2(0)+B_2^2(0)\mathrm{e}^{2\Gamma}}, \label{B2_t}
\end{equation}
where $\Gamma(t)=\int_0^t J_{\parallel}dt$.
Thus, for sufficiently large times, one obtains that $B_2\rightarrow 1$, i.e., complete reversal of the magnetization by means of a dc spin current collinear with the anisotropy axis.
From Eq. (\ref{B2_t}), it appears that the magnetic moment must be tilted away from the anisotropy axis at the initial time, i.e. $B_2(0)\neq0$, in order for the reversal process to work. This suggests a way to combine two types of ac and dc spin currents in order to shorten the reversal time. Starting with a magnetic moment oriented along $\mathbf{e}_z$, a chirped current polarized along $\mathbf{e}_x$ first tilts the moment of a certain angle with respect to the anisotropy axis (this is the technique described earlier in this work); next, a dc current polarized along $\mathbf{e}_z$ completes the reversal according to Eq. (\ref{B2_t}).

\begin{figure}[h!]
\centering
\includegraphics[width=0.8\linewidth]{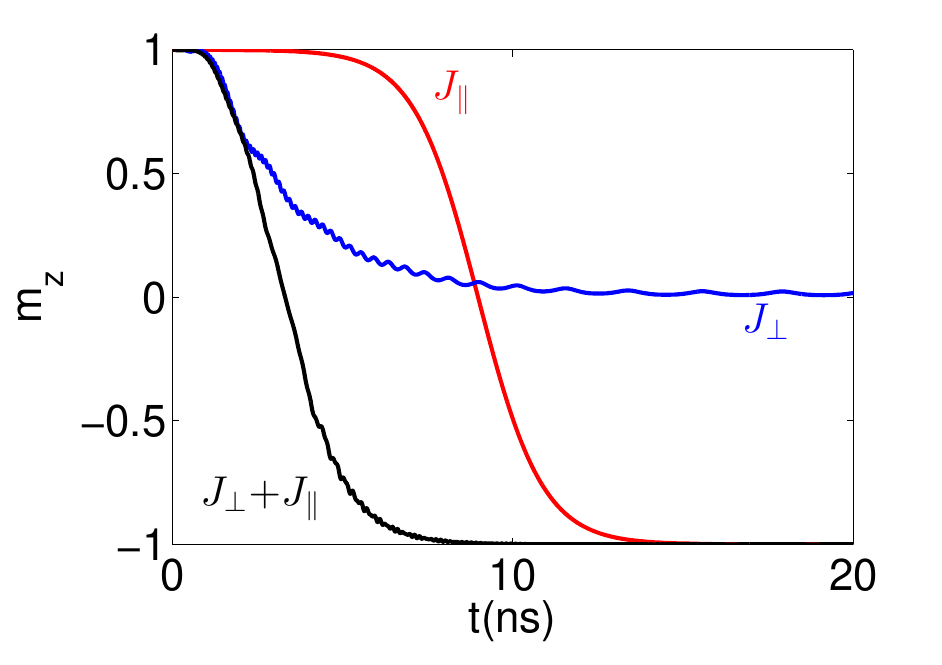}
\caption{Time evolution of $m_z$ for different types of spin currents: dc spin current of intensity $I_S=3$~mT parallel to the anisotropy axis $\mathbf{e}_z$ (red curve);
ac chirped spin current perpendicular to $\mathbf{e}_z$ with $I_S=9$~mT, $\alpha=2~\rm GHz/ns$, and $\omega_0/2\pi=20$~GHz (blue curve); and the combination of both parallel and perpendicular currents (black curve). All cases include Gilbert damping $\lambda=0.01$, but no thermal fluctuations.}
\label{fig:combo}
\end{figure}

Numerical simulations confirm this scenario (Fig. \ref{fig:combo}). Here, we show three cases where the $J_\perp$ and $J_\parallel$ currents are applied either separately or together: $J_\perp$ alone can tilt the magnetic moment only up to 90\textdegree ($m_z=0$); $J_\parallel$ alone (3~mT in this case, with an initial tilt of 1\textdegree) can reverse the magnetization completely in about 10~ns; finally, when both currents are combined, the switching time is reduced to 5~ns.
In the combined case, we used an ac spin current of magnitude 6~mT, although the theoretical threshold amplitude is close to 9~mT. This shows that the simultaneous use of the two types of excitations leads to a reduction of both the switching time and the autoresonance threshold for the $J_\perp$ component.

\section{Conclusion}
In this work we explored the potential use of chirped spin currents to manipulate and control the magnetization dynamics.
Such chirped currents could be produced by means of commercially available Arbitrary Waveform Generators, which can now reach the desired frequency range \footnote{See for instance: http://www.tek.com/signal-generator/awg5000-arbitrary-waveform-generator}.

We have shown that a chirped spin current polarized in the direction normal to the anisotropy axis can capture the magnetic moment into autoresonance and drive its precession to a stable angle (up to $90^\circ$ with respect to the anisotropy axis) on a nanosecond timescale.
The precession time (time it takes to bring the magnetization to precess at a certain angle) can also be finely controlled.
Finally, thermal noise does not alter the basic features of this scenario, and only requires a slightly larger spin current.
Thus, the autoresonant approach is particularly flexible and robust (it only requires that the spin-current frequency varies slowly, irrespective of the specific form of this variation), and should be capable of controlling with high finesse the magnetization oscillations even in very small nano-objects.

In addition, we showed that, by adding a small static magnetic field antiparallel to the anisotropy axis, it is possible to fully reverse the magnetization using a chirped spin current polarized in the direction perpendicular to the anisotropy axis.
A second method to switch the magnetization relies on the
combination of different types of spin currents.
Different scenarios that combine chirped microwave fields with ac or dc spin currents could also be envisaged \cite{finocchio2006,taniguchi2016} in the future.

{\it Acknoledgments.---}
We acknowledge the financial support of the French Agence Nationale de la
Recherche through project Equipex UNION, grant ANR-10-EQPX-52. L.F. acknowledges the support of the Israel Science Foundation. We thank Matthieu Bailleul for his careful reading of the manuscript and helpful suggestions.


\end{document}